\documentclass[aps,pra,amsmath,amssymb,superscriptaddress,notitlepage,twocolumn]{revtex4-1}
\usepackage[english]{babel}
\usepackage[utf8]{inputenc}
\usepackage[T1]{fontenc}
\usepackage[unicode]{hyperref}
\usepackage{multirow}
\usepackage{float}
\usepackage{xcolor}
\usepackage{tikz}
\usetikzlibrary{matrix}
\hypersetup{
  pdfkeywords={},
  pdfsubject={},
  pdfcreator={}}
  
\usepackage{todonotes}

\begin{document}

\author{András Bodor}
\affiliation{Department of Quantum Optics and Quantum Information, Institute for Solid State Physics and Optics, Wigner Research Centre for Physics, 1525 P.O. Box 49, Hungary}
\affiliation{Department of Differential Equations, , Budapest University of Technology and Economics, Egry József u. 1., Budapest, Hungary H-1111}
\affiliation{Institute of Physics, University of Pécs, Ifjúság útja 6., Pécs, Hungary H-7624}
\author{Orsolya Kálmán}
\affiliation{Department of Quantum Optics and Quantum Information, Institute for Solid State Physics and Optics, Wigner Research Centre for Physics, 1525 P.O. Box 49, Hungary}
\author{Mátyás Koniorczyk}
\affiliation{Department of Quantum Optics and Quantum Information, Institute for Solid State Physics and Optics, Wigner Research Centre for Physics, 1525 P.O. Box 49, Hungary}

\date{January 24, 2022}
\title{Error-free interconversion of nonlocal boxes}

\begin{abstract}
  Understanding the structure of nonlocal correlations is
  important in many fields ranging from fundamental questions of physics to device-independent cryptography. We present a
  protocol that can convert extremal two-party--two-input nonlocal
  no-signaling boxes of any type into any other extremal two-party--two-input
  nonlocal no-signaling box perfectly. Our results are exact, and even though the
  number of required boxes cannot be determined in advance, their expected
  number is finite. Our protocol is adaptive and demonstrates for the first time the usefulness
    of using no-signaling boxes in different causal orders by the
    parties.
\end{abstract}

\maketitle

\section{Introduction}\label{sec:Intro}
Nonlocality is amongst the most intriguing features of nature. Since the seminal paper of Einstein, Podolsky, and
Rosen~\cite{PhysRev.47.777}; and Bell's
quantification~\cite{PhysicsPhysiqueFizika.1.195}, the structure of
nonclassical correlations has been studied
extensively~\cite{RevModPhys.86.419}, with
implications e.g. on communication
theory~\cite{PhysRevLett.92.127901},
cryptography~\cite{RevModPhys.92.025002}, or game
theory~\cite{PhysRevA.101.062115}.

One possible way to study nonlocal correlations is to introduce a device (a so-called box) which has two separated, non-interacting parts, one at Alice, and one at Bob.
Alice chooses an input $x$ from a set of possible inputs, and receives a result $a$
from a result set. 
Similarly, Bob's input is $y$, resulting in an output $b$. 
The behavior of the box is fully described by the conditional probability distribution $p_{ab|xy}$. If $p_{ab|xy}$ does not correspond to a statistical mixture of boxes with two parts that operate independently in parallel on Alice's and Bob's side, then the box is called a nonlocal box. 

An important class of correlations is the one whose elements obey the
no-signaling condition, compatible with the theory of special
relativity. Mathematically, the no-signaling conditions can be formulated as:
\begin{eqnarray}
    \forall x_1, x_2, b, y \quad \sum_{a} p_{ab|x_1y}=\sum_{a} p_{ab|x_2y}   \nonumber\\ 
\forall y_1, y_2, a, x \quad \sum_{b} p_{ab|xy_1}=\sum_{b} p_{ab|xy_2}. 
\end{eqnarray}
These equations imply the existence of local marginals and, together with the
normalization of probabilities, define the no-signaling
polytope in the space of the conditional probabilities $p_{ab|xy}$. They are necessary and
sufficient for a box not to be useful for direct
communication~\cite{WOS:000445224900008}. In what follows, we will refer to nonlocal no-signaling boxes simply as "nonlocal boxes". 

Correlations realized by quantum systems form a convex
subset of the no-signaling polytope, which can be characterized by a
series of semidefinte
programs~\cite{1367-2630-10-7-073013,PhysRevLett.98.010401}. The
no-signaling polytope is a mathematically simpler structure which includes
supraquantum behaviors that cannot be described in the framework of quantum mechanics. Understanding the complete structure of nonlocal boxes is of fundamental importance. Notably, the extremal
points, i.e., the vertices of the polytope are of special interest.
The most frequently mentioned example is the Popescu-Rorhrlich (PR)
box~\cite{KT85,Popescu1992}, which is the extremal point of the no-signaling polytope in the two-input
two-output case. Such "maximally nonlocal" correlations would enable incredible communicational and computational power~\cite{Pawowski2009, PhysRevLett.96.250401}. On the other hand, somewhat surprisingly, they appear to underperform quantum correlations in randomness certification~\cite{PhysRevLett.114.160502}, and in correlation-assisted multiprover interactive proofs \cite{ji2021mip}. Networks of PR boxes have been used very recently to study the structure of threepartite correlations~\cite{PhysRevA.104.012210}. These examples indicate that it is possible to study nonlocality from a resource theory point of view \cite{de_Vicente_2014}. 

Regarding nonlocal correlations as a resource, it becomes important to know what kind of other correlations can be obtained if one has access to a given type of correlation. This corresponds to the question of how different nonlocal boxes can be interconverted, i.e., how boxes with certain input and output sets and behaviors can be used together to implement another box with different input and/or output sets and behavior. Barrett et al.~\cite{PhysRevA.71.022101} enumerated all extremal bipartite
nonlocal boxes with two inputs and arbitrary number of outputs. In addition, they proved that extremal two-input nonlocal boxes of a given type can be converted to any other type, with an arbitrarily small error. More precisely, they showed that $\forall \varepsilon > 0$ there exists a number of $d$-boxes $n$ so that these boxes can simulate a $d'$-box with an error probability of at most $\varepsilon$. Jones and Masanes~\cite{PhysRevA.72.052312} presented a protocol to exactly simulate
any binary-output nonlocal box with PR-boxes. Forster and Wolf~\cite{PhysRevA.84.042112} solved the general case of converting any
type of extremal nonlocal boxes to any other type, also with arbitrarily small
error. 

There is a similar concept related to manipulating nonlocal boxes, namely, nonlocal correlation distillaiton, in which the target box has the same input-output arrangement as the (not necessarily extremal) resource box~\cite{PhysRevA.80.062107,Lang_2014}. Very recently, Karvonen~\cite{PhysRevLett.127.160402} has studied the question of interconverting noncontextual and nonlocal resources in the context of generalized resource theory. In particular, he showed that the independent use of an ancillary correlated resource cannot catalyze any interconversion of correlations, which is an important structural property.

It is also pointed out in Ref.~\cite{PhysRevLett.127.160402} that \emph{adaptive} protocols, that is, when nonlocal boxes are used in a way that the input of one box can depend on another box's output, have been studied only to a limited extent thus far. 
Indeed, the no-signaling conditions allow nonlocal boxes to
  be used asynchronously by the parties. Hence, it is possible that given two boxes, Alice uses box 1 first and her input
  to box 2 depends on the output, while Bob uses box 2 first,
  and then uses box 1 with an input depending on the previous
  output he received. The possibility of such a ``crossed wiring'' is prevalently
  known (e.g. it is also mentioned in Ref.~\cite{PhysRevA.71.022101} as a side remark),
  but to our knowledge there are no protocols 
  which exploit this.

In fact, ``crossed wiring'' 
  means that the boxes are used in a different (even though definite)
  causal order by the two parties. The question of causal order is deeply related to
  separabilty~\cite{oreshkov_quantum_2012}. Indefinite causal order
  has been recognized as a resource in quantum
  communication~\cite{feix_quantum_2015, ebler_enhanced_2018} and
  computing~\cite{chiribella_quantum_2013}. It was verified
  experimentally~\cite{rubino_experimental_2017} and recently has
  also been studied in the context of general relativty.

The question of causal order has also been studied in the device-independent context, which covers supraquantum
  (including extremal no-signaling) correlations ~\cite{baumeler_device-independent_2016}. Although ``crossed
  wiring'' does not realize an indefinite causal order, it is 
  an unusual causal structure that can potentially have implications
  in this direction.

In this paper, we present a protocol which relies on the "different causal order" application of nonlocal boxes, enabling a perfect (error-free) interconversion of extremal two-input nonlocal boxes. The paper is organized as follows. In Section~\ref{subsec:Prior} we present relevant prior work on the interconversion of nonlocal boxes, then, in Sec.~\ref{subsec:errfree} our error-free protocol. In Sec.~\ref{sec:Gen} some modified versions of the error-free protocol are given with which one can extend the directly reachable range of the output boxes. We conclude in Section~\ref{sec:Sum}.

\section{Interconversion protocol}\label{sec:Main}

\subsection{Prior work} \label{subsec:Prior}
Our protocol can be considered an extension of the results of Barrett et al.~\cite{PhysRevA.71.022101}. Let us recapitulate their main results. First, they showed that every extremal nonlocal box is equivalent to a $d$-box for some
integer $d$.  In a $d$-box the input of both parties are binary: $x, y \in \left\lbrace0,1\right\rbrace$, and
the box outputs for them $a$ and $b$ with values $\{0,1,\ldots ,d-1\}$.
The nonzero $p_{ab|xy}$ probabilities are uniform (all
equal to $1/d$) for inputs and outputs which satisfy
$(b - a)\mod d= xy$, and zero otherwise. Two nonlocal boxes are considered equivalent if one can be converted into the other
by exchanging the roles of Alice and Bob, or permuting the inputs of Alice,
permuting the inputs of Bob, permuting the outputs of an input, or deleting
an input, where the output is deterministic.
Furthermore, in Ref.~\cite{PhysRevA.71.022101} three protocols were presented in order to perform interconversions between different boxes. (We note that in these algorithms $x$ and $y$ denote the inputs that Alice and Bob wish to enter into the yet-to-be simulated box.) The three protocols are the following:
\begin{description}
\item[Protocol 1] given a $d_1$ and a $d_2$-box, a $d_1 d_2$-box can
be simulated (without error).
Alice enters $x$ into the $d_1$-box. If the output $a_1$ is $d_1 -1$,
she enters $x$ into the $d_2$-box, otherwise enters $0$ to the $d_2$-box.
Her overall output is computed as $a_2 d_1 + a_1$.
Bob enters $y$ into both boxes. His overall output is computed as $b_2 d_1 + b_1$.
\item[Protocol 2] given a $d_1 d_2$-box, a $d_1$-box can
be simulated (without error). 
Both parties enter their original
input into the $d_1 d_2$-box, and take the output modulo $d_1$.
\item[Protocol 3] given $n$ pieces of a $d_1$-box, a $d_2$-box can
be simulated provided that $d_2\leqslant d_1^n$ (with arbitrarily small error by increasing $n$). Alice and Bob simulate a $d_1^n$-box using Protocol 1, and take the
output modulo $d_2$. Note that this protocol is not error free: although the zero probabilities remain zero, the nonzero probabilities will deviate a little from the uniform
distribution.
\end{description}

With the help of Protocols 1-3 of Ref.~[\onlinecite{PhysRevA.71.022101}] any $d$-box can be simulated  using $d'$-boxes, however, there is still a large class of boxes for which a nonzero error is unavoidable. This is the case for incommensurable $d$ and $d'$. 

\subsection{Error-free interconversion protocol}\label{subsec:errfree}
In what follows, we will show that one can construct a protocol which can operate withour error.
Our protocol requires a specific causal order in the use of the
boxes: the parties have to use certain boxes in opposite
order, so that the inputs on the box used later depends on the output
of the box that is used first. We assume that a nonlocal box can be queried only once, so that e.g. when the actions of the parties are repeated twice, we assume the use of two boxes of the same type, and not to query the same box twice. We speak of the "number of boxes" in this sense. 

\textbf{Lemma 1:} \textit{Given two $d$-boxes Alice and Bob can convert them into one $(d+1)$-box with probability $(d^2-1)/d^2$ or get a specific output on both sides which signifies an unsuccessful conversion attempt, and this happens with probability $1/d^2$.}

The conversion can be carried out using a single round of Protocol 4 below. Before introducing the protocol and proving Lemma 1, let us state our main result first:

\textbf{Theorem 1:} \textit{Given an infinite supply of $d$-boxes Alice and Bob can realize one $(d+1)$-box with probability $1$ and the expected number of actually consumed boxes is $2 d^2/(d^2-1)$.}

\textbf{Proof of Theorem 1:} Repeating the rounds of Protocol 4 will eventually lead to success. As the probability that
the protocol does not halt in the current round is $1/d^2$, the
expected number of rounds can be computed by summing the series $\left(1-1/d^2\right)\sum_{k=1}^\infty k(1/d^2)^{k-1}$.

\textbf{Protocol 4:}
In a single round the parties consume two $d$-boxes.   
Alice inputs $x$ (the value which would be the input of the box to be simulated) into the \emph{first} box. 
If the result is $0$, she inputs
$0$ to the second box, otherwise she inputs $x$ to the second box
as well. If the overall result is not $00$, then Alice terminates the protocol and the output is $a_1$ if $a_1\leqslant a_2$, and $a_1+1$
if $a_1>a_2$. If her overall result is $00$, then she starts a new round repeating these steps using
two fresh $d$-boxes. On the other side, 
Bob inputs $y$ to the \emph{second} box (note the inverted causal order as compared to Alice's side, i.e. the ``crossed wiring''). If the result is $0$, he inputs
$0$ to the first box, otherwise he inputs $y$ to the first box
as well. If the overall result is not $00$, then Bob terminates the protocol and his output is $b_1$ if
$b_1\leqslant b_2$,
and $b_1+1$ if $b_1>b_2$. If his overall result is $00$ then he starts a new round using two new $d$-boxes (similarly to Alice's procedure). 

We note that since the $00$ result can only be obtained by Alice and Bob in coincidence in the same round there is no need for them to communicate classically in order to start a new round of the protocol.

\begin{table}[ht]
  $$
  \begin{array}{|lll||cccc|cccc|}
    \hline
    x\downarrow& y\rightarrow&&\multicolumn{4}{|c}{0}&  \multicolumn{4}{c|}{1}\\
    &\quad & in &
       00 & 00 & 00 & 00 & 01 & 11 & 01 & 11\\
    & & \downarrow & \downarrow
       & \downarrow & \downarrow & \downarrow & \downarrow & \downarrow & \downarrow & \downarrow \\
    & in\rightarrow out & out &
       00 & 01 & 10 & 11 & \bf{00} & \bf{01} & \bf{10} & \bf{11}\\
    \hline
    \hline
    \multirow{4}{*}{0}& 00\rightarrow 00 & &  1/4 & 0 & 0 & 0 &  1/4 & 0 & 0 & 0 \\
    \cline{2-3}& 00\rightarrow 01 &&          0 & 1/4 & 0 & 0 &  0 & 1/4 & 0 & 0 \\
    \cline{2-3}& 00\rightarrow 10 &&          0 & 0 & 1/4 & 0 &  0 & 0 & 1/4 & 0 \\
    \cline{2-3}& 00\rightarrow 11 &&          0 & 0 & 0 & 1/4 &  0 & 0 & 0 & 1/4 \\
    \hline
    \multirow{4}{*}{1}& 10\rightarrow \bf{00} & &  1/4 & 0 & 0 & 0 &  \color{blue}{1/4} & 0 & 0 & 0 \\
    \cline{2-3}& 10\rightarrow \bf{01} &&          0 & 1/4 & 0 & 0 &  0 & 0 & 0 & \color{red}{1/4} \\
    \cline{2-3}& 11\rightarrow \bf{10} &&      0 & 0 & 1/4 & 0 &  0 & \color{brown}{1/4} & 0 & 0 \\
    \cline{2-3}& 11\rightarrow \bf{11} &&      0 & 0 & 0 & 1/4 &  0 & 0 & \color{green}{1/4} & 0 \\
    \hline
  \end{array}
  $$
  \caption{The joint probabilities pertaining to the inputs and outputs of two 2-boxes
utilized according to Protocol 4. The order within the bit pairs corresponds to the temporal order of the boxes on Alice's side. The probabilities are determined by multiplying the
probabilities of the individual 2-boxes, which are given as $p_{a_{i}b_{i}|x_{i}y_{i}}=1/2$
if $b_{i}\oplus a_{i}=x_{i}y_{i}$ and $0$ otherwise, where the subscripts $i=1,2$ refer
to the first and second box, respectively. The notion of the colors and output pairs displayed in boldface is explained in the caption of Table~\ref{tab:22arrows}.}
  \label{tab:22to3}
\end{table}

Before proving the correctness of Protocol 4 for arbitrary $d$, let us illustrate,
as a simple example, how it can convert two 2-boxes (PR-boxes)
into a 3-box. The joint probabilities for a single round of the protocol are presented
in Table \ref{tab:22to3}. 
The inputs $xy$ for the target box divides the table into four blocks.
Although there are 16 possible $\left[\text{in},\text{out}\right]$ combinations for each user
in each block, we tabulate only
those four that appear in the protocol with nonzero probability.
Additionally in each $4\times 4$ block there is only a single nonzero entry in each
row and column. For instance, in the $(x,y)=(1,1)$ block the probability pertaining to row 3
and column 3 means that Bob had entered 1 to the second box, received 0, then entered 0 to the
first box and received 1. Meanwhile, Alice had entered 1 to the first box, received 1,
therefore entered 1 also to the second box, and received 0. Because of the inputs,
the outputs of the first box should be correlated ($x_{1} + y_{1} \mod 2=0$) and those of the second box
should be anticorrelated ($x_{2}+y_{2} \mod 2=1$). But both outputs are correlated ($a_{1}=b_{1}$,
$a_{2}=b_{2}$), therefore this case is impossible, so the matrix entry is 0.
Observe that the upper left entries of the blocks (corresponding to the outputs $00$ for
both parties) always have probability $1/4$. According to the protocol this is the
indication that the round fails and must be repeated. As the $00$ outputs can only occur
in coincidence, both parties recognize this failure without the need for any communication.
The remaining lower right $3\times3$ submatrices of the blocks in the table are equivalent
to a $3$-box (presented in Table \ref{tab:3box}), by the following relabelling of the outputs (independent of the inputs):
$01\rightarrow 0$, $10\rightarrow 2$, $11\rightarrow 1$. Overall, each round of the protocol
succeeds with
probability $3/4$ and fails otherwise.

\begin{table}[ht]
  $$
  \begin{array}{|ll||ccc|ccc|}
    \hline
    x\downarrow& y\rightarrow&\multicolumn{3}{|c}{0}&  \multicolumn{3}{c|}{1}\\    
      & a\downarrow b\rightarrow & 0 & 1 & 2 & 0 & 1 & 2 \\    
    \hline
    \hline
    \multirow{3}{*}{0}& 0 &  1/3 & 0 & 0 &  1/3 & 0 & 0 \\
    \cline{2-2} & 1 &          0 & 1/3 & 0 &  0 & 1/3 & 0 \\
    \cline{2-2}& 2 &           0 & 0 & 1/3 &  0 & 0 & 1/3 \\
    \hline
    \multirow{3}{*}{1}& 0 &  1/3 & 0 & 0 & 0 & 1/3 & 0 \\
    \cline{2-2}& 1 &         0 & 1/3 & 0 & 0 & 0 & 1/3 \\
    \cline{2-2}& 2 &         0 & 0 & 1/3 & 1/3 &  0 & 0 \\
    \hline
  \end{array}
  $$
  \caption{The conditional probability distribution of the targeted 3-box.}
  \label{tab:3box}
\end{table}

 The equivalence with the $3$-box can also be seen by ordering the possible
nonzero-probability outputs into a table, as shown in Table \ref{tab:22arrows}. This
representation reveals that -- apart from a trivial one-cycle (00) -- there exists
a length-3 cycle among the nontrivial output pairs. The relabelling of the outputs
is then straightforward.

\begin{table}[ht]
\begin{tikzpicture}
\matrix (m) [matrix of math nodes,row sep=10, column sep = 10]
{
  01 & 11 \\
  00 & 10 \\
};
\draw[->,thick,draw=red] (m-1-1.east) -- (m-1-2.west);
\draw[->,thick,draw=green] (m-1-2.south) -- (m-2-2.north);
\draw[->,thick,draw=brown] (m-2-2.north west) -- (m-1-1.south east);
\draw[->,thick,draw=blue] (m-2-1.west) arc [radius = 0.5em, start angle=90,end angle=360];
\end{tikzpicture}
  \caption{Nontrivial output pairs of the parties and their interdependence when applying
Protocol 4 on two boxes with $d=2$. Pairs of numbers represent output pairs $a_1\,a_2$ or
$b_1\,b_2$ when $x=y=1$ (highlighted in bold in Table \ref{tab:22to3}).  If there is a nonzero
probability for a given pair of outputs, then an arrow is drawn from one pair to the other in
the sense that Alice's output stands at the base, while Bob's output stands at the point
of the arrow. The colors correspond to the colored entries of Table \ref{tab:22to3}.
Due to the nature of the $d$-boxes, the correlations are unique
i.e. $a_1\,a_2$ determines $b_1\,b_2$ and vice-versa. Therefore there is only one arrow
starting and ending at every point.}
\label{tab:22arrows}
\end{table}

\textbf{Proof of Lemma 1:} The proof can be accomplished by analyzing Protocol 4 for general inputs. If any of $x$ or $y$
is $0$, then Alice and Bob receive identical outputs, so the matrix in
these blocks is proportional to the identity matrix, therefore it is sufficient to
analyze the $x=y=1$ case only. As both parties make deterministic steps,
the $d^2\times d^2$ matrix of possible outputs will still have only $d^2$
nonzero entries, one in each row and each column, so we can consider it
as a permutation matrix, similar to the one in Table \ref{tab:22arrows}.
Studying the the cycle structure of this permutation matrix reveals the type of boxes hiding in the result.
The conditional probability distribution of any $d$-box can be transformed to a form similar to that in Table~\ref{tab:22arrows}, in which 3 blocks are $1/d$ times the identity matrix, and the fourth block
is $1/d$ times a permutation matrix with a single $d$-cycle (cyclic shift of each
element one step to the right).
As the different cycles of the joint probability distributions of a single round of the protocol divide the blocks into different submatrices each containing a cycle with a certain length, after receiving their outputs
Alice and Bob can identify the respective submatrix without communication. If they find that their output does not correspond to the desired submatrix, they can start a new round and can eventually reach the targeted submatrix: the one which simulates the $d^\prime$-box (with $d^\prime$ equals the length of the cycle in this submatrix). Thus the cycle
structure determines the types of boxes that can be simulated by this protocol.

The permutation matrix is a permutation of the set with
elements of the form $(c_1,c_2)$, where $0\leqslant c_i < d$ are integers.
If $(a_1,a_2),(b_1,b_2)$ is a possible simultaneos output of Alice and
Bob, then the permutation corresponding to the matrix takes the element
$(a_1,a_2)$ to $(b_1,b_2)$. (Note, that for every output of Alice, $(a_1,a_2)$
there is exactly one possible output of Bob $(b_1,b_2)$).

There are 4 (2 $\times$ 2) cases:
\begin{enumerate}
\item If $a_1=b_2=0$, then $a_2=b_2$, and $a_1=b_1$, therefore
the element $(0,0)$ goes to $(0,0)$.
\item If $a_1=0$ and $b_2\neq0$, then $a_2=b_2$, and $(a_1+1)\mod d=b_1$, therefore
the elements of the form $(0,a_2)$ go to $(1,a_2)$ ($a_2\neq 0$).
\item If $a_1\neq0$ and $b_2=0$, then $(a_2+1)\mod d=b_2$, and $a_1=b_1$, therefore
the elements of the form $(a_1,d-1)$ go to $(a_1,0)$ (where $a_1\neq 0$).
\item If $a_1\neq0$ and $b_2\neq0$, then $(a_2+1)\mod d=b_2$, and $(a_1+1) \mod d=b_1$,
therefore
the elements of the form $(a_1,a_2)$ go to $(a_1+1 \mod d,a_2+1 \mod d)$
(where $a_1\neq 0$ and $a_2\neq d-1$).
\end{enumerate}
In order to obtain the cycle structure, let us examine the orbits of the elements $(0,a_2)$.
If $a_2=0$, then $(0,0)$ does not move, the orbit has one element, thus, this is
a 1-cycle.
Otherwise, $(0,a_2)$ first moves to $(1,a_2)$, then $(1+s,a_2+s)$ for
$1\leqslant s \leqslant d-a_2-1$, then to $(d-a_2,0)$ then to
$(d-a_2+s,s)$ for
$1\leqslant s \leqslant a_2$, then returns to $(0,a_2)$. These are $d+1$
steps altogether, which is a $(d+1)$-cycle. The cycles do not overlap because
the first coordinate cannot become $0$ before returning, therefore
we get $d-1$ $(d+1)$-cycles, and these cases cover all pairs, because $d^2=(d-1)(d+1)+1$.
It is easy to see that the output function labels the consecutive elements of every
$(d+1)$-cycle from $0$ to $d$. This completes the proof of Lemma 1.

\begin{table}[ht]
\begin{tikzpicture}
\matrix (m) [matrix of math nodes,row sep=10, column sep = 10]
{
  & \phantom{00} & \phantom{00} & \phantom{00} & \phantom{00} & \phantom{00} & \phantom{00}\\
  \phantom{00} & 04 & 14 & 24 & 34 & 44 & \phantom{00}\\
  \phantom{00} & 03 & 13 & 23 & 33 & 43 & \phantom{00}\\
  \phantom{00} & 02 & 12 & 22 & 32 & 42 & \phantom{00}\\
  \phantom{00} & 01 & 11 & 21 & 31 & 41 & \phantom{00}\\
  \phantom{00} & 00 & 10 & 20 & 30 & 40 & \phantom{00}\\
  & \phantom{00} & \phantom{00} & \phantom{00} & \phantom{00} & \phantom{00} & \phantom{00}\\
};
\draw[->,draw=blue, thick] (m-2-2.east) -- (m-2-3.west);
\draw[->,draw=blue, thick] (m-2-3.north) -- (m-1-3.south);
\draw[->, thick] (m-2-4.north) -- (m-1-4.south);
\draw[->,draw=red, thick] (m-2-5.north) -- (m-1-5.south);
\draw[->,draw=green, thick] (m-2-6.north) -- (m-1-6.south);
\draw[->,draw=blue, thick] (m-3-1.north east) -- (m-2-2.south west);
\draw[->, thick] (m-3-2.east) -- (m-3-3.west);
\draw[->, thick] (m-3-3.north east) -- (m-2-4.south west);
\draw[->,draw=red, thick] (m-3-4.north east) -- (m-2-5.south west);
\draw[->,draw=green, thick] (m-3-5.north east) -- (m-2-6.south west);
\draw[->,draw=blue, thick] (m-3-6.north east) -- (m-2-7.south west);
\draw[->, thick] (m-4-1.north east) -- (m-3-2.south west);
\draw[->,draw=red, thick] (m-4-2.east) -- (m-4-3.west);
\draw[->,draw=red, thick] (m-4-3.north east) -- (m-3-4.south west);
\draw[->,draw=green, thick] (m-4-4.north east) -- (m-3-5.south west);
\draw[->,draw=blue, thick] (m-4-5.north east) -- (m-3-6.south west);
\draw[->, thick] (m-4-6.north east) -- (m-3-7.south west);
\draw[->,draw=red, thick] (m-5-1.north east) -- (m-4-2.south west);
\draw[->,draw=green, thick] (m-5-2.east) -- (m-5-3.west);
\draw[->,draw=green, thick] (m-5-3.north east) -- (m-4-4.south west);
\draw[->,draw=blue, thick] (m-5-4.north east) -- (m-4-5.south west);
\draw[->, thick] (m-5-5.north east) -- (m-4-6.south west);
\draw[->,draw=red, thick] (m-5-6.north east) -- (m-4-7.south west);
\draw[->,draw=green, thick] (m-6-1.north east) -- (m-5-2.south west);
\draw[->,draw=blue, thick] (m-6-3.north east) -- (m-5-4.south west);
\draw[->, thick] (m-6-4.north east) -- (m-5-5.south west);
\draw[->,draw=red, thick] (m-6-5.north east) -- (m-5-6.south west);
\draw[->,draw=green, thick] (m-6-6.north east) -- (m-5-7.south west);
\draw[->,draw=blue, thick] (m-7-3.north) -- (m-6-3.south);
\draw[->, thick] (m-7-4.north) -- (m-6-4.south);
\draw[->,draw=red, thick] (m-7-5.north) -- (m-6-5.south);
\draw[->,draw=green, thick] (m-7-6.north) -- (m-6-6.south);
\draw[->] (m-6-2.west) arc [radius = 0.5em, start angle=90,end angle=360];
\end{tikzpicture}
  \caption{Nontrivial output pairs of the parties and their interdependence when applying Protocol 4 on two boxes with $d=5$. Pairs of numbers represent output pairs $a_1\,a_2$ or $b_1\,b_2$.  If there is a correlation between a given pair of outputs, then an arrow is drawn from one pair to the other in the sense that Alice's output stands at the base, while Bob's output stands at the point of the arrow. The 4 possible six-cycles are highlighted in red, green, blue and black.
           }
  \label{tab:55arrows}
\end{table}

As an illustration of the proof, the case $d=5$ is displayed in Table
\ref{tab:55arrows}. If all four outputs are different from 0, then the
outputs of both boxes differ by one, so in the most part of the
matrix, the arrow is upward diagonal. If, however, the first output of
Alice is one, then the second outputs must coincide, therefore
horizontal arrows start from the first column, and similarly, vertical
arrows from the first row.  Alice and Bob can get $00$ at the same
time, therfore a circular arrow is drawn into the corresponding
cell. The cycles are moving mainly diagonally, but at the first column
they jump one position to the left, and at the first row, jump
back. So nearly all diagonals correspond to some six-cycle, while some
disappeared like in the so-called ''vanishing leprechaun''
puzzle~\cite{leprechaun,leprechaun2}.  As can be seen, there is a one-cycle ($00$)
and 4 six-cycles, so a single round converts two $5$-boxes into a
$6$-box with probability $24/25$, and is unsuccessful with probability $1/25$. Alice and Bob can unambiguously identify this latter case, and continue with the protocol.

\begin{table}[tbh]
\begin{tikzpicture}
\matrix (m) [matrix of math nodes,row sep=10, column sep = 10]
{
  & \phantom{00} & \phantom{00} & \phantom{00} & \phantom{00} & \phantom{00} & \phantom{00}\\
  \phantom{00} & 04 & 14 & 24 & 34 & 44 & \phantom{00}\\
  \phantom{00} & 03 & 13 & 23 & 33 & 43 & \phantom{00}\\
  \phantom{00} & 02 & 12 & 22 & 32 & 42 & \phantom{00}\\
  \phantom{00} & 01 & 11 & 21 & 31 & 41 & \phantom{00}\\
  \phantom{00} & 00 & 10 & 20 & 30 & 40 & \phantom{00}\\
  & \phantom{00} & \phantom{00} & \phantom{00} & \phantom{00} & \phantom{00} & \phantom{00}\\
};
\draw[->, thick] (m-2-2.east) -- (m-2-3.west);
\draw[->, thick] (m-2-3.east) -- (m-2-4.west);
\draw[->, thick] (m-2-4.north) -- (m-1-4.south);
\draw[->, thick] (m-2-5.north) -- (m-1-5.south);
\draw[->, thick] (m-2-6.north) -- (m-1-6.south);
\draw[->, thick] (m-3-1.north east) -- (m-2-2.south west);
\draw[->, thick] (m-3-2.east) -- (m-3-3.west);
\draw[->, thick] (m-3-3.east) -- (m-3-4.west);
\draw[->, thick] (m-3-4.north east) -- (m-2-5.south west);
\draw[->, thick] (m-3-5.north east) -- (m-2-6.south west);
\draw[->, thick] (m-3-6.north east) -- (m-2-7.south west);
\draw[->, thick] (m-4-1.north east) -- (m-3-2.south west);
\draw[->, thick] (m-4-2.east) -- (m-4-3.west);
\draw[->, thick] (m-4-3.east) -- (m-4-4.west);
\draw[->, thick] (m-4-4.north east) -- (m-3-5.south west);
\draw[->, thick] (m-4-5.north east) -- (m-3-6.south west);
\draw[->, thick] (m-4-6.north east) -- (m-3-7.south west);
\draw[->, thick] (m-5-1.north east) -- (m-4-2.south west);
\draw[->, thick] (m-5-2.east) -- (m-5-3.west);
\draw[->, thick] (m-5-3.east) -- (m-5-4.west);
\draw[->, thick] (m-5-4.north east) -- (m-4-5.south west);
\draw[->, thick] (m-5-5.north east) -- (m-4-6.south west);
\draw[->, thick] (m-5-6.north east) -- (m-4-7.south west);
\draw[->, thick] (m-6-1.north east) -- (m-5-2.south west);
\draw[->, thick] (m-6-4.north east) -- (m-5-5.south west);
\draw[->, thick] (m-6-5.north east) -- (m-5-6.south west);
\draw[->, thick] (m-6-6.north east) -- (m-5-7.south west);
\draw[->, thick] (m-7-4.north) -- (m-6-4.south);
\draw[->, thick] (m-7-5.north) -- (m-6-5.south);
\draw[->, thick] (m-7-6.north) -- (m-6-6.south);
\draw[->] (m-6-2.west) arc [radius = 0.5em, start angle=90,end angle=360];
\draw[->] (m-6-3.west) arc [radius = 0.5em, start angle=90,end angle=360];
\end{tikzpicture}
  \caption{Nontrivial output pairs of the parties and their interdependence when applying the modified version of Protocol 4 in the case of $d=5$. 
           }
  \label{tab:55arrows2}
\end{table}

\section{Generalizations}\label{sec:Gen}
Protocol 4 can be slightly modified to simulate other boxes as well, not only $d+1$-ones. To achieve this one needs to change the number of cases when Alice or Bob
enters $0$ into their respective ``second'' box. The case when Bob
enters $0$ to the first box as a result of getting either $0$ or $1$ as the output of the second box is 
illustrated in Table \ref{tab:55arrows2}: the diagonally moving
cycle jumps left once and right twice, so after traversing the matrix once
it restarts at the next diagonal, covering the whole matrix (except for the two
stationary points).
Thus, it contains two one-cycles, 
and a single 23-cycle. In general, one can say that, if one party enters $0$ in one case
and the other one enters $0$ in two cases, then a $d$-box can be converted
into a $(d^2-2)$-box. The success probability of a single round is $(d^2-2)/d^2$.

Another possible modification is when both parties enter $0$ to their respective ``second box'' if their ``first'' output is smaller than some value $s$, i.e., if  $a_1<s$, $b_2<s$ (where $1\leqslant s < d$). In this case they can get a ($d+s$)-box, but with decreasing success
probability. The permutation then has $s^2$ one-cycles ($a_1<s$ and $b_2<s$)
while the $(d+s)$-cycles contain four diffrerent sections: (i) $(0,a_2)$ moves to $(q_1,a_2)$, where $1\leqslant q_1 \leqslant s$, 
(ii) $(s+q_2,a_2+q_2)$, where $1\leqslant q_2 \leqslant d-a_2-1$,
(iii) $(d-a_2-1,q_3)$, where $1\leqslant q_3 \leqslant s$, and finally
(iv) $(d-a_2-1+q_4,s+q_4)$, where $1\leqslant q_4 \leqslant d-s-1$.
Thus, the success probability of a single round is $(d^2-s^2)/d^2$.
As a simple example one can choose $d=5$ and $s=4$, in which case a single round can simulate a $9$-box with $9/25$ (see Table~\ref{tab:55arrows44}).

\section{Conclusions}\label{sec:Sum}
We have presented a protocol (and its relevant modified versions),
which, together with Protocols 1 and 2 of
Ref.[\onlinecite{PhysRevA.71.022101}], enable the conversion of any $d$-boxes
into any other $d^\prime$-box without error. 
In the other similar protocols known so far
the parties have to agree on the number of turns to go below a fixed
error, and they need to communicate if they want to further improve on
it. Our protocol, on the other hand, allows for unlimited number of iterations in principle, with a halting condition that can be verified without communication, and an error-free conversion. The expected number of required iterations is finite.
There may be other possibilities to
modify our protocol, such as, combining two boxes of different size.

To the best of our knowledge, our conversion protocol is the first
one to utilize the fact that Alice and Bob are allowed to query their
parts of the boxes in \emph{different causal order}. It is an open
question whether there exists a protocol for realizing error-free interconversion of nonlocal boxes without ``crossed wiring''.  It would be also interesting to find a useful
protocol in which there are three boxes involved, and the order in
which certain boxes are used depends on the output of some other
boxes. This could potentially demonstrate the use of indefinite causal
order in the present device-independent context.

\begin{table}[t]
\begin{tikzpicture}
\matrix (m) [matrix of math nodes,row sep=10, column sep = 10]
{
& \phantom{00} & \phantom{00} & \phantom{00} & \phantom{00} &
\phantom{00} & \phantom{00}\\
\phantom{00} & 04 & 14 & 24 & 34 & 44 & \phantom{00}\\
\phantom{00} & 03 & 13 & 23 & 33 & 43 & \phantom{00}\\
\phantom{00} & 02 & 12 & 22 & 32 & 42 & \phantom{00}\\
\phantom{00} & 01 & 11 & 21 & 31 & 41 & \phantom{00}\\
\phantom{00} & 00 & 10 & 20 & 30 & 40 & \phantom{00}\\
& \phantom{00} & \phantom{00} & \phantom{00} & \phantom{00} &
\phantom{00} & \phantom{00}\\
};
\draw[->] (m-3-1.north east) -- (m-2-2.south west);
\draw[->] (m-2-2.east) -- (m-2-3.west);
\draw[->] (m-2-3.east) -- (m-2-4.west);
\draw[->] (m-2-4.east) -- (m-2-5.west);
\draw[->] (m-2-5.east) -- (m-2-6.west);
\draw[->] (m-2-6.north) -- (m-1-6.south);
\draw[->] (m-3-2.west) arc [radius = 0.5em, start angle=90,end angle=360];
\draw[->] (m-3-3.west) arc [radius = 0.5em, start angle=90,end angle=360];
\draw[->] (m-3-4.west) arc [radius = 0.5em, start angle=90,end angle=360];
\draw[->] (m-3-5.west) arc [radius = 0.5em, start angle=90,end angle=360];
\draw[->] (m-3-6.north east) -- (m-2-7.south west);
\draw[->] (m-4-2.west) arc [radius = 0.5em, start angle=90,end angle=360];
\draw[->] (m-4-3.west) arc [radius = 0.5em, start angle=90,end angle=360];
\draw[->] (m-4-4.west) arc [radius = 0.5em, start angle=90,end angle=360];
\draw[->] (m-4-5.west) arc [radius = 0.5em, start angle=90,end angle=360];
\draw[->] (m-4-6.north) -- (m-3-6.south);
\draw[->] (m-5-2.west) arc [radius = 0.5em, start angle=90,end angle=360];
\draw[->] (m-5-3.west) arc [radius = 0.5em, start angle=90,end angle=360];
\draw[->] (m-5-4.west) arc [radius = 0.5em, start angle=90,end angle=360];
\draw[->] (m-5-5.west) arc [radius = 0.5em, start angle=90,end angle=360];
\draw[->] (m-5-6.north) -- (m-4-6.south);
\draw[->] (m-6-2.west) arc [radius = 0.5em, start angle=90,end angle=360];
\draw[->] (m-6-3.west) arc [radius = 0.5em, start angle=90,end angle=360];
\draw[->] (m-6-4.west) arc [radius = 0.5em, start angle=90,end angle=360];
\draw[->] (m-6-5.west) arc [radius = 0.5em, start angle=90,end angle=360];
\draw[->] (m-6-6.north) -- (m-5-6.south);
\draw[->] (m-7-6.north) -- (m-6-6.south);
%
angle=390];
\end{tikzpicture}
  \caption{Nontrivial output pairs of the parties and their interdependence when applying the second modified version of Protocol 4 in the case of $d=5$ and $s=4$. 
           }
  \label{tab:55arrows44}
\end{table}

\begin{acknowledgements}
  This research was supported by the National Research, Development,
  and Innovation Office of Hungary under project numbers K133882 and
  K124351 and the Quantum Information National Laboratory of Hungary. We thank Tam\'as
  Vértesi, Tam\'as Kiss, Tam\'as Geszti, Lajos Di\'osi for useful
  discussions and the revision of the manuscript.
\end{acknowledgements}


\end{document}